# Semantic Query Optimisation with Ontology Simulation

Siddharth Gupta[1] and Narina Thakur[2]

[1]UG Research Scholar, [2]Asst. Professor
[1,2]Department of Computer Science, Bharati Vidyapeeth's College of Engineering, New Delhi, India
siddharth_bvcoe@hotmail.com
narinat@gmail.com

***ABSTRACT***

*Semantic Web is, without a doubt, gaining momentum in both industry and academia. The word "Semantic" refers to "meaning" – a semantic web is a web of meaning. In this fast changing and result oriented practical world, gone are the days where an individual had to struggle for finding information on the Internet where knowledge management was the major issue. The semantic web has a vision of linking, integrating and analysing data from various data sources and forming a new information stream, hence a web of databases connected with each other and machines interacting with other machines to yield results which are user oriented and accurate. With the emergence of Semantic Web framework the naïve approach of searching information on the syntactic web is cliché. This paper proposes an optimised semantic searching of keywords exemplified by simulation an ontology of Indian universities with a proposed algorithm which ramifies the effective semantic retrieval of information which is easy to access and time saving.*

## 1. INTRODUCTION

The World Wide Web is a congregation of billion web pages which are adhered to each other through hyperlinks. In 1989 Tim Barners-Lee [1] invented the World Wide Web and marked the advent of Web 1.0 in which all the web pages were published with no user interaction. Web 2.0 based on equal user participation, collaboration; inter personal connectivity, interconnected applications [4]. Some of the applications are blogging, Facebook, Flickr, MySpace, Google, and YouTube where both producer and consumers can interact with each other [3]. The major drawback of Web 2.0 is its lack of interpretability between machines. Because of the lack of metadata and knowledge management crisis, powerful and complex algorithms are required by the search engines in order to parse the keywords requested by the user. The future web, semantic web is based on the principal of interoperability between machines and giving them power to think [5], aims at attaching metadata, specifying relations between web resources and knowledge management, in order to process and integrate data by the users. Hence semantic web is a web of databases and not of documents, queried by SPARQL [12]. RDF attaches metadata specify relations between the resources based on XML. Ontologies are another major semantic web technology built above RDF aims at providing strong semantics and vocabulary [6]. They link the data on the basis of logical reasoning, common vocabulary and analytical thinking. This paper elucidates searching of information based on semantic understanding of words with simulation of Ontology of Indian Universities using protégé is a tool. An algorithm based on semantic searching of information is exemplified by an example. The main idea behind it is to collectively manage all the related information of Indian universities on a single platform which aggregates the information related to conferences, inter-college competitions, exchange of ideas, interaction among the students hence knowledge management and sharing on a single platform.
The Semantic Web is an extension of the World Wide Web with new technologies and standards that enable interpretation and processing of data and useful information for extraction by a computer[15]. The Web contains a huge amount of data but computers alone cannot understand or make any decisions with this data. The solution is the Semantic Web. The Semantic Web is not a separate Web but an extension of the World Wide Web, in which information is given well defined meaning, better enabling computers and people to work in cooperation. [9] Sometimes it is said that the semantic web will make data become “smart”. What would it mean, for data to be smart? Smart data means that the web of information becomes so richly interconnected that it can become much smarter than humans. [10]
With the emergence of Semantic Web framework the naïve approach of searching information on the syntactic web is cliché. The World Wide Web is a congregation of billion web pages which are adhered to each other through hyperlinks. Many internet users daily activity is web search only and these users end up in an endless quench to retrieve relevant information pertaining to the user in shortest possible time A Web 2.0 based search engine is having major drawback as it lacks interpretability between machines, metadata and knowledge management crisis [4]. Powerful and complex algorithms are required by the search engines in order to parse the keywords requested by the user. The future web, semantic web is based on the principal of interoperability between machines and giving them power to think [3][7], aims at attaching metadata, specifying relations between web resources and knowledge management, in order to process and integrate data by the users.
For e.g. Let a user types a keyword set X. Then by looking at the texts found by means of X, some other words related to X can be determined like antonyms and synonyms. Obviously, these texts may not include any query word and they include merely one related word or more [1]. This aims at finding the probabilities of the keywords with the maximum likelihood occurrence if keywords in a set. This method which will help us in assigning priority to the web pages based on the high probability.

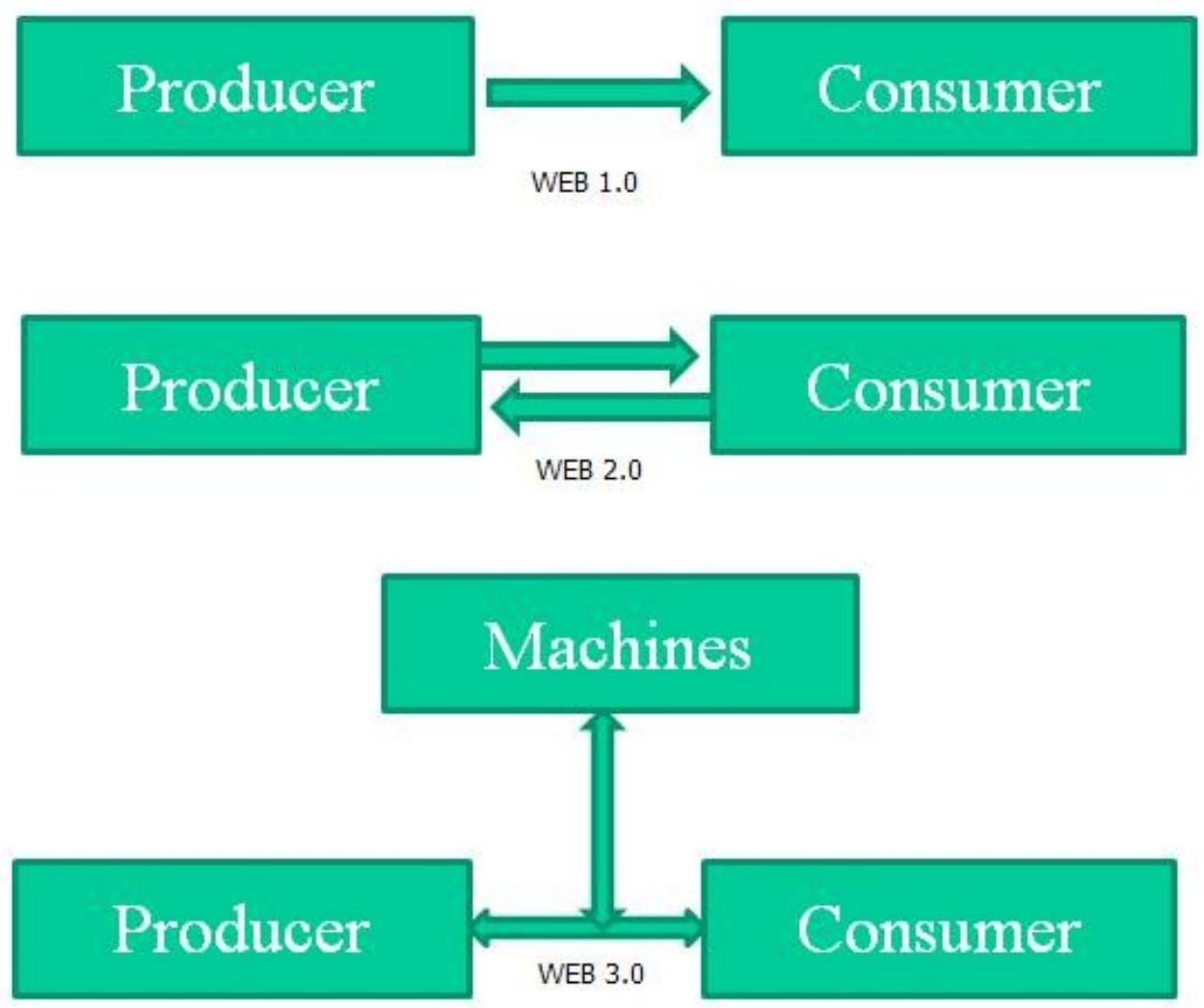

Fig.1 Evolution of WEB 3.0

This paper discusses them in detail with simulation of ontology of Indian universities using protégé which shows how the sematic web works on the whole.

## 2. SEMANTIC WEB ARCHITECTURE

The first layer URI/IRI (Uniform Resource Identifier / International Resource Identifier) is a string of standardized form in order to uniquely identify resources/documents. URL (Uniform Resource Locator) is a subset of URI. IRI is an internationally accepted form. XML layer with the XML namespace and XML schema ensures that there is common syntax used in the semantic web. XML is the key for platform independence and exchange data using a common language[13][15]

The core format for representation of data in semantic web is RDF which is based on triples (subject – predicate – object) and forms a graph pertaining to the given data in the same form [7]. It is the grammar of the document whereas XML is the words understood by machines. OWL (Web ontology Language) uses description logics and provides strong semantics. For querying RDF and OWL ontologies, SPARQL is used which acts as a query tool similar to SQL [8]. The correct logics with the rules implied proves the ontologies which then intertwined with trusted inputs to yield trusted outputs. Digital signatures and cryptography is applied in order to maintain security while information exchange.

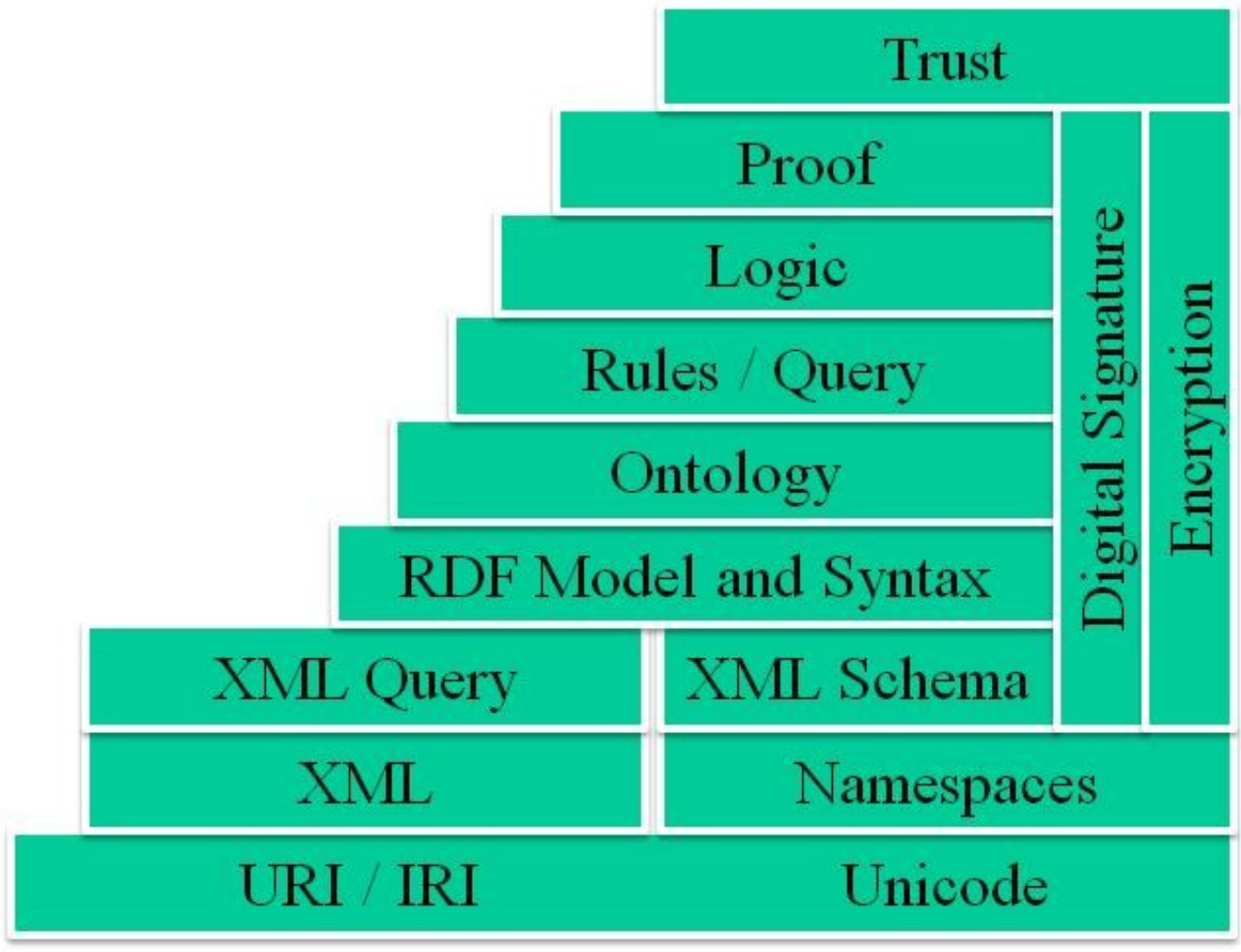

Fig 2. Semantic web architecture [7].

***RDF***

Resource Descriptive Framework (RDF) is the backbone of semantic web, which is the standard model for data interchange on the web. It is linked with the other resources via URI which uniquely identifies the documents. It works on the XML technology hence making data in each database understood by each other in XML form and hence platform independent. The RDF model is based on the idea of triple. It breaks down the data into three parts namely resource, property and property-value which are nothing but subject, predicate and object in simple grammar [7] [8].

For example:

Sentence 1: Bharti Vidyapeeth has a student, Siddharth Gupta

Here, Bharti Vidyapeeth is the subject (resource), student is the predicate (property) and Siddharth Gupta is the object (property-value).

Subject: http://bvcoend.ac.in/info#BhartiVidyapeeth

Predicate: http://bvcoend.ac.in/info#student

Object: Siddharth Gupta

Sentence 2: Siddharth Gupta has a nickname Sid

Here, Siddharth Gupta is the subject (resource), nickname is the predicate (property) and Sid is the object (property-value).

where, http://bvcoend.ac.in/info# is the URI which uniquely identifies the resources. Standard URI for RDF is
http://www.w3.org/1999/02/22-rdf-syntax-ns#

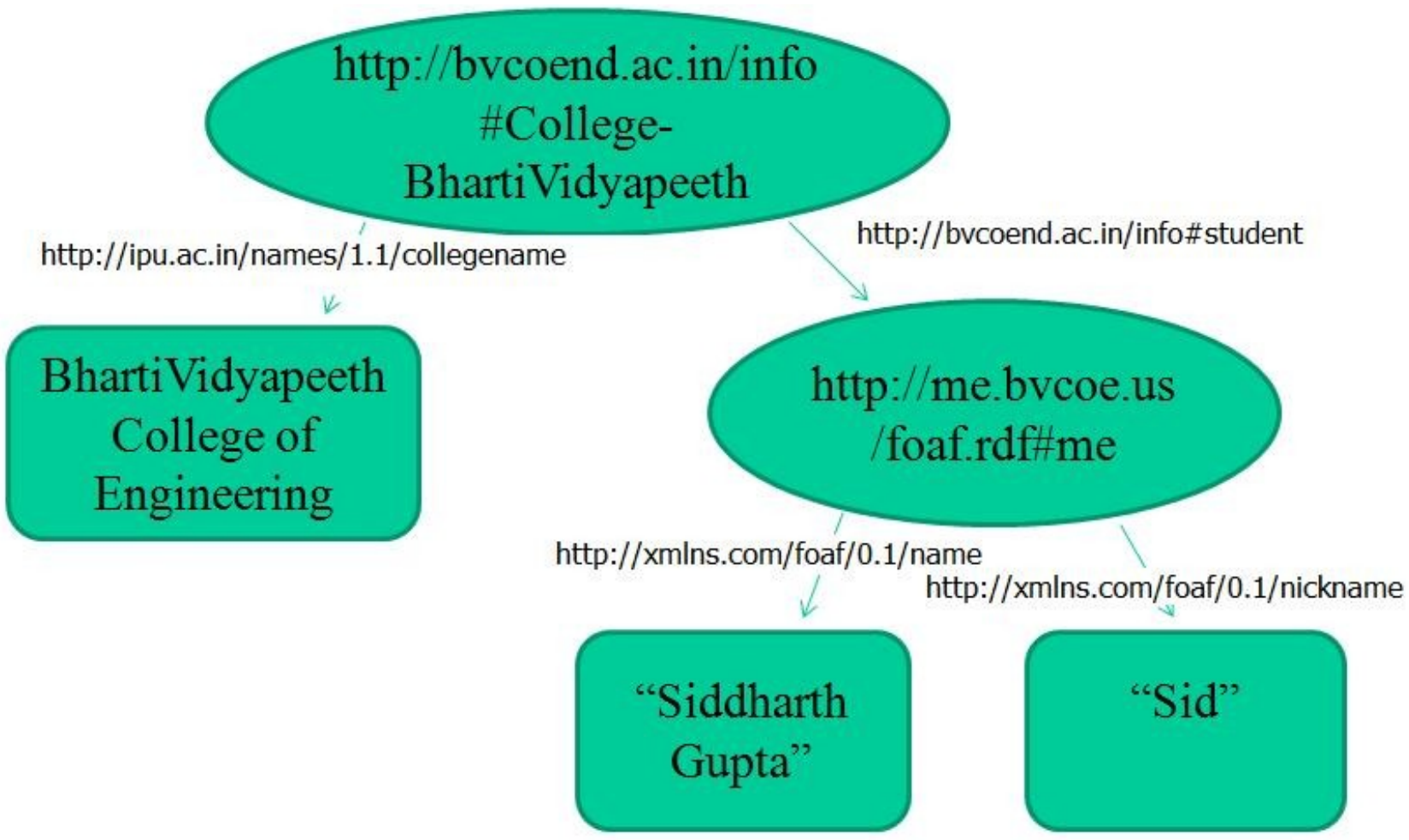

Fig 3. RDF graph tree

The graph shows a FOAF (Friend of a Friend) who describes Siddharth Gupta. We examine the graph is in triplets of subject – predicate – object. There is an other language called RDFS (RDF schema) which uses vocabulary to define resources.  Other triple formats are: N3, turtle, N-Triples. The XML is hence generated depending on the graph.

<rdf:RDF
xmlns:rdf="http://www.w3.org/1999/02/22-rdf-syntax-ns#"
xmlns:info="http://www.bvcoend.ac.in/info#"
xmlns:foaf="http://xmlns.com/foaf/0.1/">
<rdf:Description rdf:about=
"http://www.bvcoend.ac.in/info#College-"BhartiVidyapeeth">
<info:student rdf:resource=
"http://me.bvcoend.us/foaf.rdf#me"/>
</rdf:Description>
</rdf:RDF>

***OWL***

OWL is a stronger language as compared to RDF and has much greater machine interpretability. Thomas Gruber defines it as "explicit specification of conceptualisation" [9], which are the relationships that concepts, objects and other entities hold with each other. According to Barry smith "the ontology is a classification of entities, represented by nodes in a hierarchal tree" [10].

The ontologies are derived form the real world conceptualisation shared by humans as a knowldege base and implemented in digital described through machine readable languages such as OWL, XML. Everything is derived from <owl:Thing> under which we have the classes, properties and instances. The main component of ontology is Class which describes concepts in the domain. Secondly are the properties which are further classified into object properties and datatype properties which describes various features and attributes of the concepts and then individual instances of the classes are made which in together makes a knowledge-base.
The results plotted are based on logical reasoning and linking

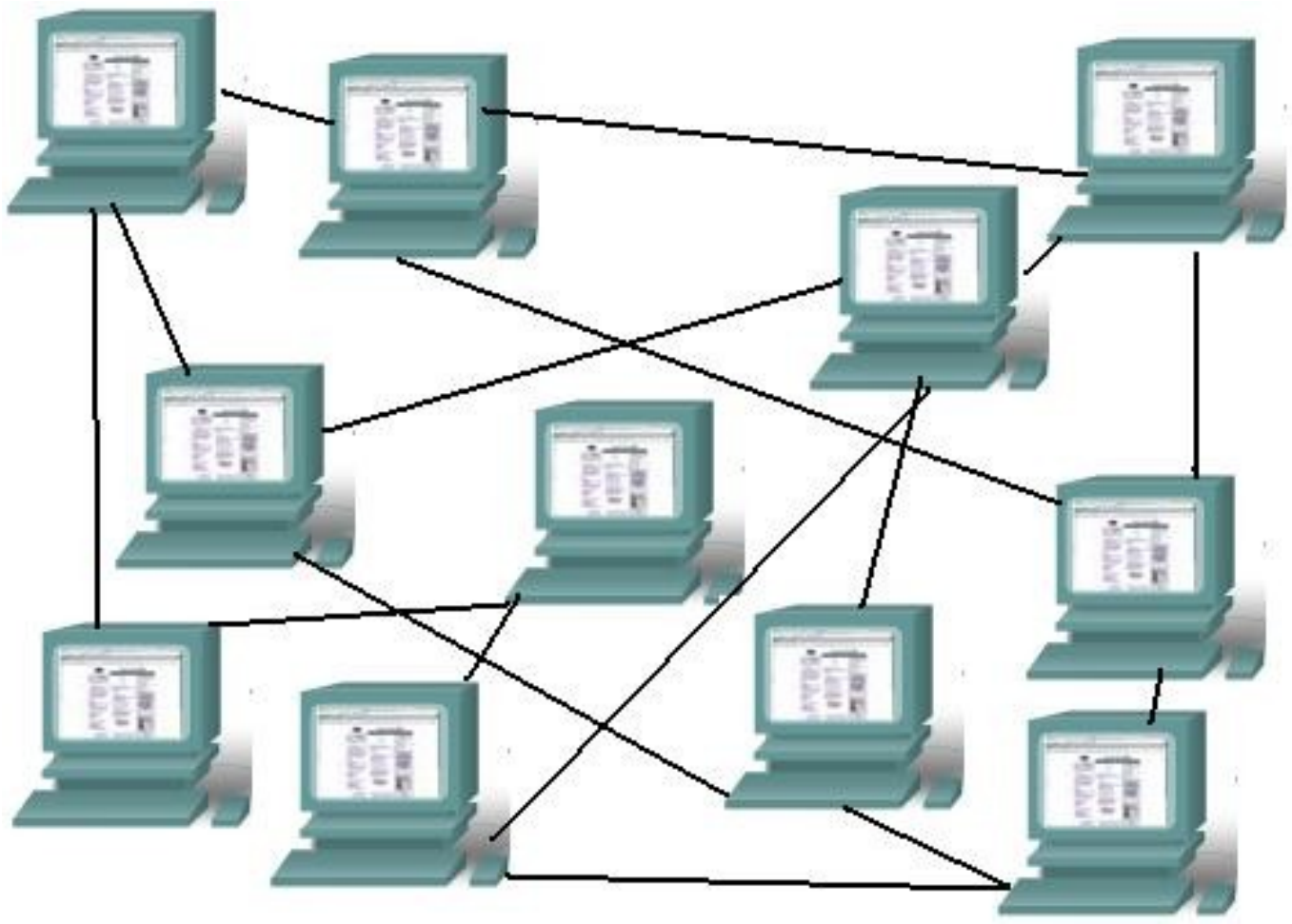

Fig 4. Interconnection between machines exchanging data

***SPARQL***

SPARQL (Simple Protocol and RDF Query Language) is similar to SQL and used to access native graph based RDF stores and extract data from the traditional databases, hence yielding perfect results.

An example of a SELECT query follows:

```
PREFIX foaf:   <http://xmlns.com/foaf/0.1/>
   SELECT ?college ?university
   WHERE { ?name foaf:college ?college .
         ?name foaf:universtity ?university . }
```

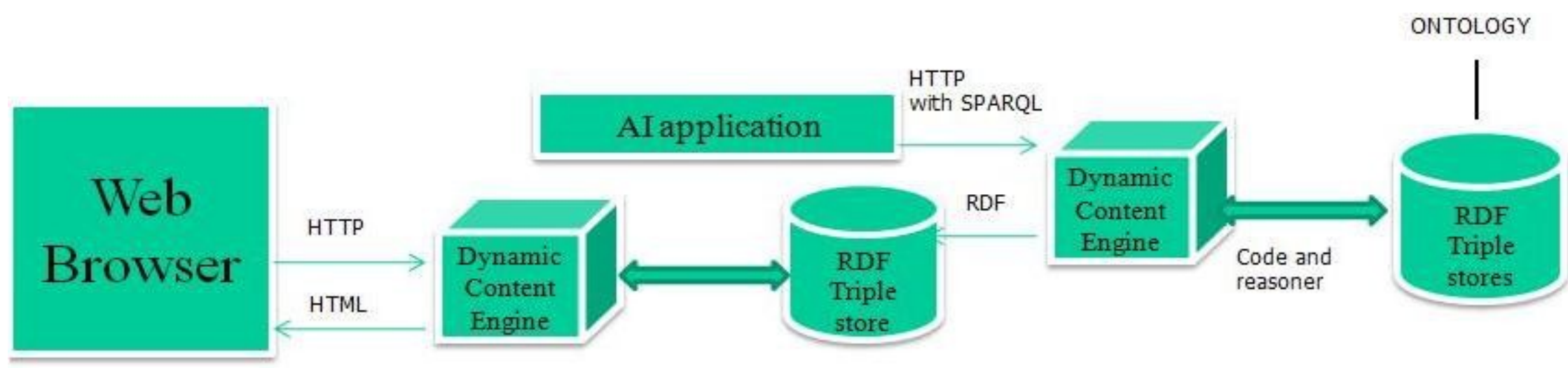

Fig 5. Querying RDF stores with SPARQL

## 3. ONTOLOGY OF INDIAN UNIVERSITIES

Using protégé [11], which is a free, open source platform for constructing, visualizing and manipulating domain models and knowledge based applications with ontologies we have simulated ontology of Indian Universities. It implies that if different Indian colleges websites providing information about various events, services, information are published and shared under same ontology in terms they all use, then computer agents can share and aggregate information from these websites and hence provide a more viable solution to the user query about Indian universities/ colleges. The domains of the ontology defined can be reused by other ontology, thus integrating several existing ontologies under one large ontology describing a large domain.

The name of my ontology representing unique URI:

http://www.indianuniversities/ourontology1.owl

Defining Classes:

We define three main classes namely Universities, Courses and States. The University class has been further subclassed into Colleges. The instances of these classes will be defined later.
The courses class contains a list of all the courses as its subclasses and states contain a list of states.
All the classes are subclassed under one class <owl:Thing> which is the root of all the classes.

Defining properties:

The properties of the classes are hence defined which includes Object properties and Datatype properties.

TABLE 1

| **Object Properties** | **Domain** |
|---|---|
| | |
| hasChiefMinister | State |
| hasColleges | University |
| hasCourses | University |
| hasPrincipal | Colleges |
| hasViceChancellor | University |

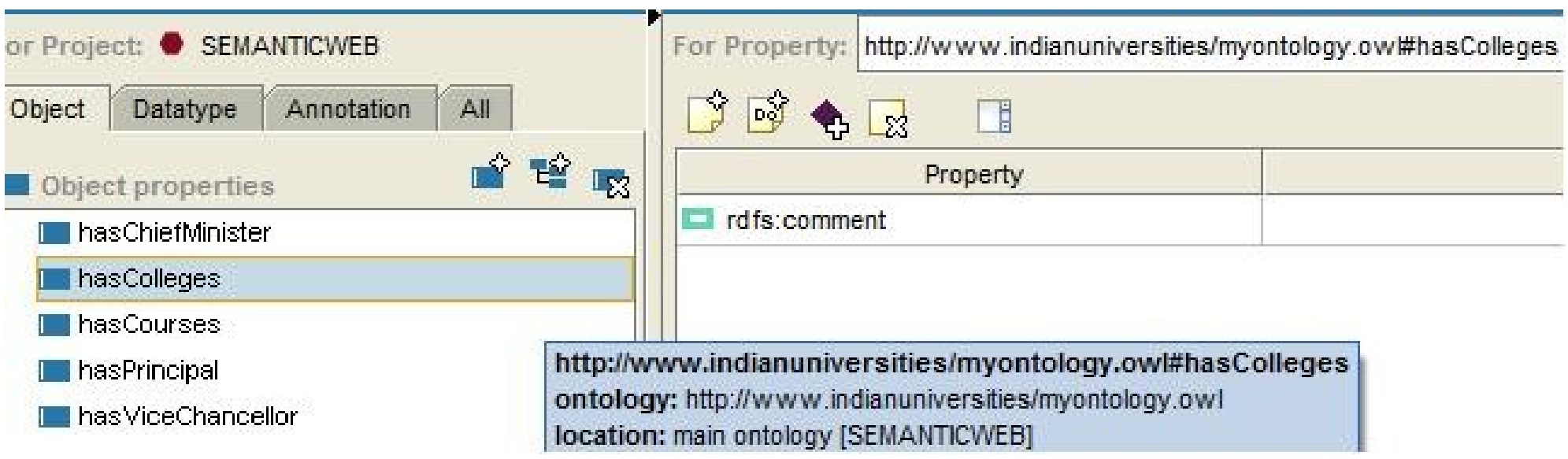

Fig 6. Object Properties

And now the domain of each property is defined to show which class they are pertaining to. Example: the domain for Datatype property hasCollegeName is Colleges and domain of Object property hasColleges is University.

TABLE 2

| Datatype Properties | Domain |
|---|---|
| | |
| hasUniversityName | Universities |
| hasCollegeName | Colleges |
| hasCourseName | Courses |
| hasStateName | States |
| hasPhoneNumber | Universities |

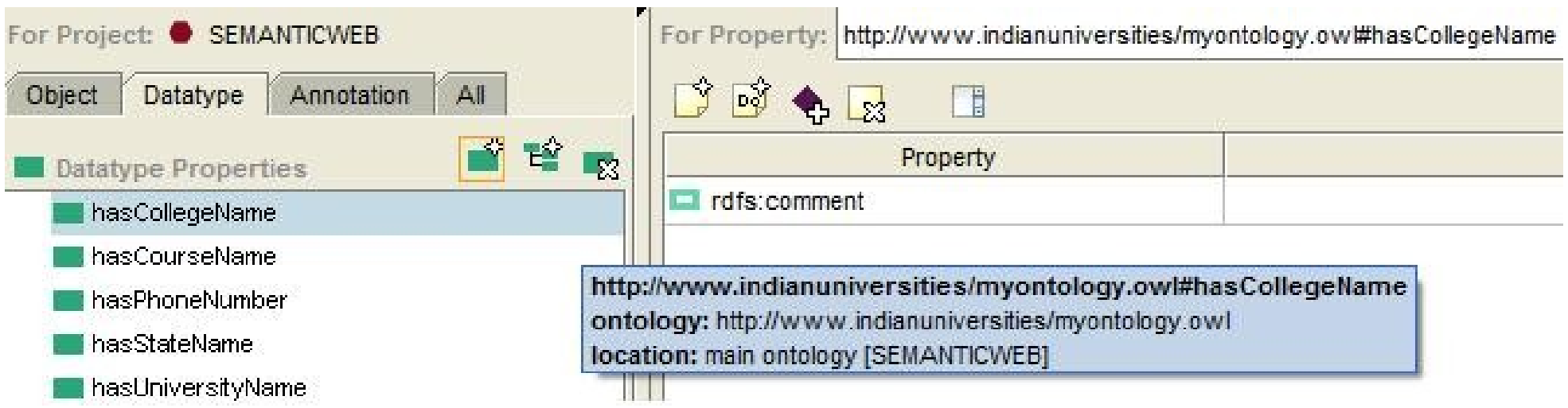

Fig 7. Datatype Properties

The assertions are made using assertion browser and new instances of the classes can be created with the details according to the properties defined.

The following subclasses were made and its instances using instance browser, which shows a column wise creation of assertions of the classes:

TABLE 3

| Universities | Colleges | Courses | States |
|---|---|---|---|
| | | | |
| IP University | BVCOE | BTECH | New Delhi |
| Delhi University | DCE | MTECH | Mumbai |
| IIT | IIT Chennai | MBA | Chennai |
| | IIT Delhi | MCA | Kolkata |
| | IIT Mumbai | MEDICAL | |
| | NSIT | | |
| | MAIT | | |

The results are hence plotted on the graphs using OWLViz tool and following simulations were done which shows the interconnection between the classes and the domain defined.
The fig 4. Shows the class hierarchy which shows the base class as <owl:Thing> and all the classes subclassed under it.

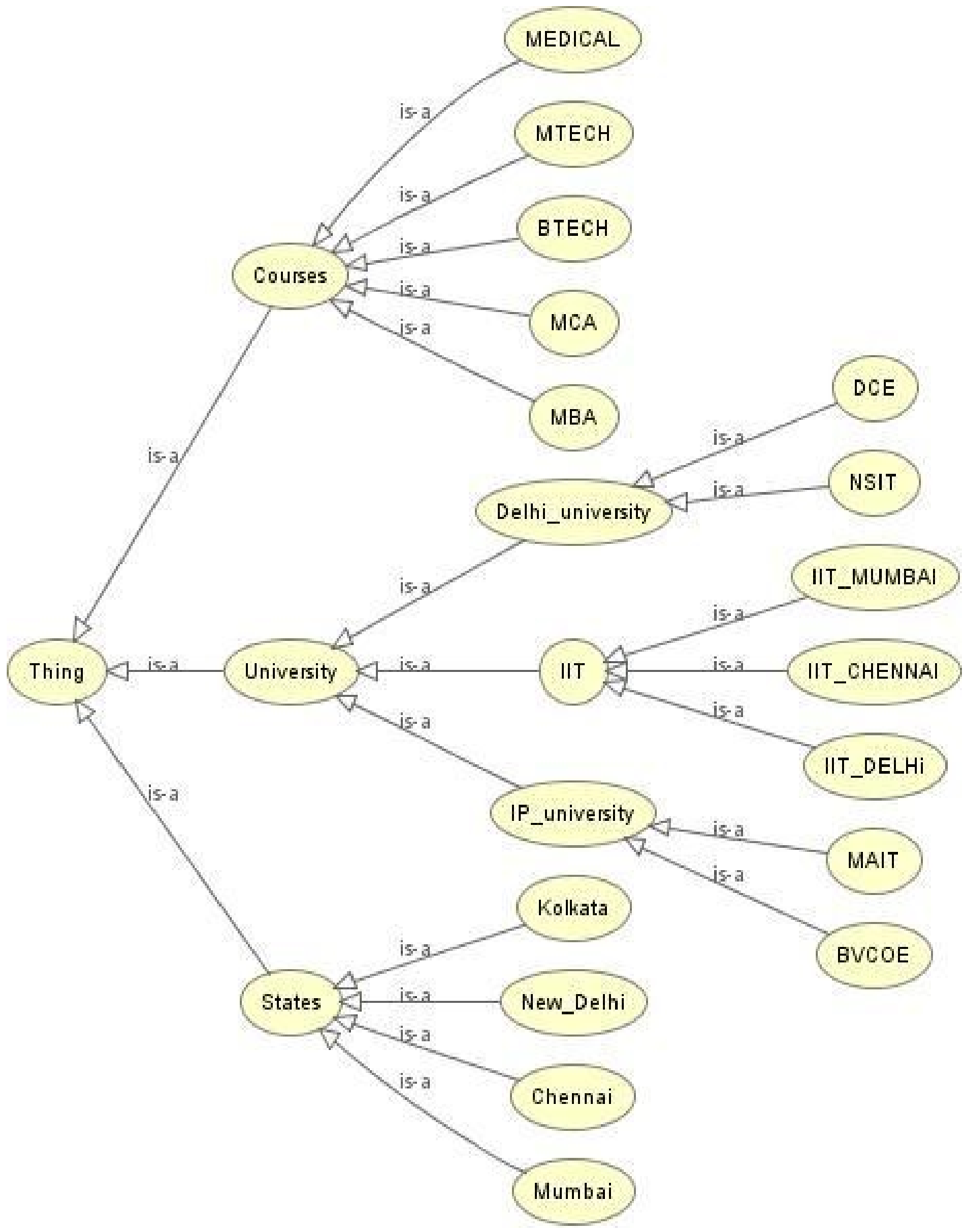

Fig 8. Class hierarchy

The figure 5 shows the classes and the individual tree of the classes and its instances in a radial layout. Hence we can see the instances of the Universities class and has a subclass Colleges which has its own set of instances and how they are interconnected with the State and Courses classes all under one root <owl:Thing>.

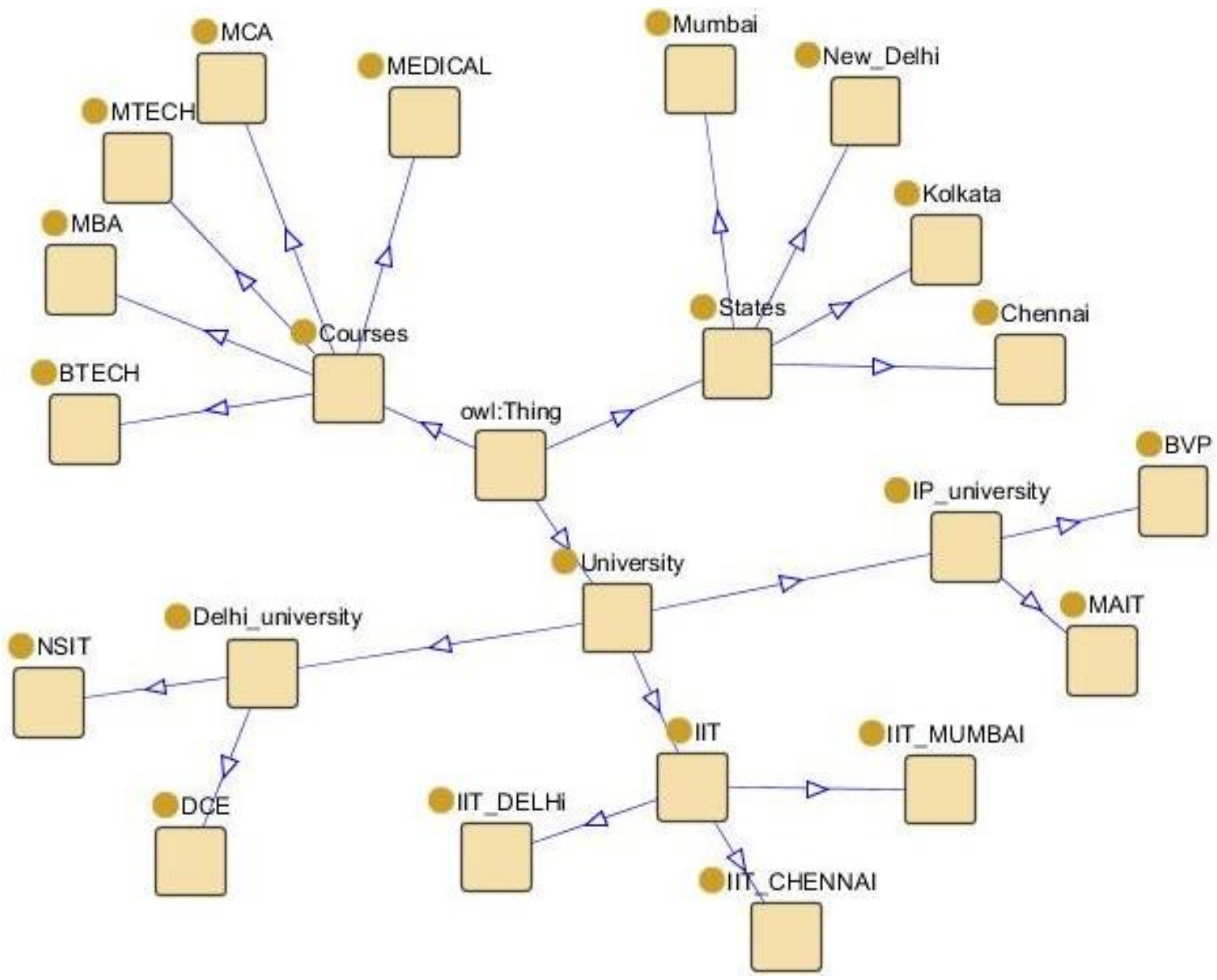

Fig 9 Radial Layout

*Algorithm ( semantic search )*

**Input : Search := { $x_1, x_2, x_3, x_4 \ldots X_n$ }**
**Output : Set O := { semantic retrieval }**
**Procedure :**
**Let Set I := Φ , Set P := Φ**
**for every keyword set $Search_i$ in web database**

**if Search ∩ $Search_i \neq \Phi$ then add $Search_i$ to I;**
**Let I := {$search_1$, $search_2$, $search_3$… $search_n$}**
**P:= $search_1 \cup search_1 \cup search_1 \cup \ldots search_n$**
**C := count number of keywords in Set P**

**O($P_i$) :={ Occurrence, Prioritise Keysets with the probability of occurrence }**

The above algorithm illustrates the semantic searching of words on the web. As we know present scenario of searching depends on heavy parsing algorithms in order to yield results as computer cannot understand the meaning of the documents. Here we have retrieved all the keysets related to the search query in the database and the union of all the keysets are taken in another set[14]. The probability of occurrence was found and according to that priority was given to it as a perfect search result.

Example :

Search Query= { "Indian Universities" }

Let Keyword sets obtained were :

$Search_1$ = { "Indian", "Courses", "AIU", "Universities", "UGC", "Top", "Colleges", "States"}
$Search_2$ = { "Indian", "UGC","AIU", "List", "Exams", "Top", "Universities"}
$Search_3$ = { "Universities", "UGC", "Colleges", "Top", "Indian", "Ranking" }
$Search_4$ = { "Indian", "States", "AIU", "Universities", "Questions", "Ranking "}
$Search_5$ = { "Indian", "Courses", "Ranking", "UGC", "States", "Universities"}

Now we have a set P in which union of all the keysets are taken

P = { "Indian", "Universities", "Courses", "AIU", "UGC", "Top", "Colleges", "States", "List", "Exams", "Ranking", "Questions" }
Now as we see Indian Universities occur in every search hence the comparison is made with reference to these two keywords.

**Occurrence ($P_i$) = {Φ, Φ, 2, 3, 4, 3, 2, 2, 1, 1, 3, 1 }**

According to the observation probability of UGC is more than other keywords hence, it is given higher priority and then AIU. Though computer doesn't understand the meaning, but we can assign priorities in order to yield semantic search as keywords with higher priority occurs in greater number of keysets and implies that it is closely related to the search query. The order of occurrence on the basis of priority is : UGC ,AIU, Top, Ranking, Courses, Colleges, states, List, Exams, Questions.

Now, on a standard google search of Indian Universities yields app.8,820,000 web pages. As we know, most of them are irrelevant to the user, hence according to our problem we can infer that if we discard the relevancy of web pages containing the keywords – List, Exams and Questions because of the lowest assigned priorities found above, we can discard almost 15% of the web pages consisting of these keywords i.e 7,497,000 pages can be considered and further refined according the assigned priorities and hence making the search more efficient.

## 4. FUTURE SCORE AND CONCLUSION

This paper focuses on how the future web might look like and the interaction between the sources of information to yield perfect and real time results with unique power of intelligence by interpreting the best possible solution for the query. Ontologies are the bases of semantic web and can be further expanded. An example ontology was simulated using protégé and results were analysed. The keywords were given priority and bases on that we have discarded certain percentage of web pages hence making the search more compact and efficient. The main idea is to collaborate the search for Indian Universities to be more informative and provide it an intellicense in order to retrieve user oriented results. The future scope of this method is that Firstly, the current ontology can be integrated into large domains resulting in expansion of knowledge base. Secondly, Information of all the universities and their corresponding colleges will be assorted under one ontology hence interoperability as well as interpretability. Thirdly,
The web will be a collection of databases with a common vocabulary for exchanging and interpreting information between machines.